\documentclass[submission,copyright,creativecommons,sharealike]{eptcs}
\providecommand{\event}{ACT 2021} %
\providecommand{\thisvolume}[1]{this volume of EPTCS, Open Publishing Association}
\usepackage{breakurl}             %
\usepackage{underscore}           %

\title{Polynomial Life: the Structure of Adaptive Systems}
\def\titlerunning{Polynomial Life}

\author{Toby St. Clere Smithe
  \institute{Topos Institute}
  \email{toby@topos.institute}
}
\def\authorrunning{T. St. Clere Smithe}

\date{\today}

\usepackage{amsfonts}
\usepackage{amssymb}
\usepackage{amsthm}
\usepackage{amsmath}

\usepackage[inline]{enumitem}
\setlist{itemsep=0em, topsep=0em, parsep=0em}
\setlist[enumerate]{label=(\alph*)}
\usepackage{mathtools} %

\usepackage{tikz}
\usepackage{tikzit}

\newcommand*{\secref}[1]{\S\ref{#1}}

\usepackage[numbers,square]{natbib}

\def\mdash{{\hbox{-}}}

\makeatletter
\newcommand{\adjunction}{\@ifstar\named@adjunction\normal@adjunction}
\newcommand{\normal@adjunction}[4]{%
  #1\colon #2%
  \mathrel{\vcenter{%
    \offinterlineskip\m@th
    \ialign{%
      \hfil$##$\hfil\cr
      \longrightharpoonup\cr
      \noalign{\kern-.3ex}
      \smallbot\cr
      \longleftharpoondown\cr
    }%
  }}%
  #3 \noloc #4%
}
\newcommand{\named@adjunction}[4]{%
  #2%
  \mathrel{\vcenter{%
    \offinterlineskip\m@th
    \ialign{%
      \hfil$##$\hfil\cr
      \scriptstyle#1\cr
      \noalign{\kern.1ex}
      \longrightharpoonup\cr
      \noalign{\kern-.3ex}
      \smallbot\cr
      \longleftharpoondown\cr
      \scriptstyle#4\cr
    }%
  }}%
  #3%
}
\newcommand{\longrightharpoonup}{\relbar\joinrel\rightharpoonup}
\newcommand{\longleftharpoondown}{\leftharpoondown\joinrel\relbar}
\newcommand\noloc{%
  \nobreak
  \mspace{6mu plus 1mu}
  {:}
  \nonscript\mkern-\thinmuskip
  \mathpunct{}
  \mspace{2mu}
}
\newcommand{\smallbot}{%
  \begingroup\setlength\unitlength{.15em}%
  \begin{picture}(1,1)
  \roundcap
  \polyline(0,0)(1,0)
  \polyline(0.5,0)(0.5,1)
  \end{picture}%
  \endgroup
}
\makeatother

\let\op=\relax

\DeclareMathOperator{\op}{^\text{op}}

\newcommand{\nn}{{\mathbb{N}}}
\newcommand{\rr}{{\mathbb{R}}}
\newcommand{\Tt}{{\mathbb{T}}}

\newcommand{\Cat}[1]{\mathbf{#1}}
\newcommand{\cat}[1]{\mathcal{#1}}
\newcommand{\Fun}[1]{\mathsf{#1}}

\newcommand{\Kl}{\mathcal{K}\mspace{-2mu}\ell}

\DeclareMathOperator*{\E}{\mathbb{E}}

\renewcommand{\d}{\mathrm{d}}

\DeclareMathOperator{\Fa}{\mathcal{F}}

\DeclareMathOperator{\Pa}{\mathcal{P}}
\DeclareMathOperator{\Qa}{\mathcal{Q}}

\newcommand{\Giry}{\mathcal{G}}

\DeclareMathOperator{\im}{\mathsf{im}}

\DeclareMathOperator{\id}{\mathsf{id}}

\DeclareMathOperator{\Set}{\Cat{Set}}

\newcommand{\xto}[1]{\xrightarrow{#1}}

\makeatletter
\newcommand{\mathoverlap}[2]{\mathpalette\mathoverlap@{{#1}{#2}}}
\newcommand{\mathoverlap@}[2]{\mathoverlap@@{#1}#2}
\newcommand{\mathoverlap@@}[3]{\ooalign{$\m@th#1#2$\crcr\hidewidth$\m@th#1#3$\hidewidth}}
\makeatother

\newcommand{\klcirc}{\bullet} %
\newcommand*{\smallklcirc}{\raisebox{0.18ex}{\scalebox{0.66}{$\klcirc$}}}
\newcommand{\klto}{\mathbin{\mathoverlap{\rightarrow}{\smallklcirc\,}}}

\newcommand{\xklto}[1]{\mathbin{\mathoverlap{\xrightarrow{#1}}{\smallklcirc\,}}}

\newcommand{\lenscirc}{
  \mathbin{\mathoverlap{\circ}{\raisebox{0.375ex}{\scalebox{1.0}[0.33]{$|$}}}}
}
\newcommand{\lensto}{\mathrel{\ooalign{\hfil$\mapstochar\mkern5mu$\hfil\cr$\to$\cr}}}

\makeatletter
\providecommand*{\xmapstofill@}{%
  \arrowfill@{\mapstochar\relbar}\relbar\rightarrow
}
\providecommand*{\xmapsto}[2][]{%
  \ext@arrow 0395\xmapstofill@{#1}{#2}%
}

\makeatletter
\def\slashedarrowfill@#1#2#3#4#5{%
  $\m@th\thickmuskip0mu\medmuskip\thickmuskip\thinmuskip\thickmuskip
   \relax#5#1\mkern-7mu%
   \cleaders\hbox{$#5\mkern-2mu#2\mkern-2mu$}\hfill
   \mathclap{#3}\mathclap{#2}%
   \cleaders\hbox{$#5\mkern-2mu#2\mkern-2mu$}\hfill
   \mkern-7mu#4$%
}
\def\rightslashedarrowfill@{%
  \slashedarrowfill@\relbar\relbar\mapstochar\rightarrow}
\newcommand\xslashedrightarrow[2][]{%
  \ext@arrow 0055{\rightslashedarrowfill@}{#1}{#2}}
\makeatother

\theoremstyle{definition}
\newtheorem{defn}{Definition}[section]

\newtheorem{ex}[defn]{Example}

\newtheorem{rmk}[defn]{Remark}
\newtheorem{obs}[defn]{Observation}

\newtheorem{prop}[defn]{Proposition}

\newtheorem{lemma}[defn]{Lemma}
\newtheorem{thm}[defn]{Theorem}

\newtheorem*{thm*}{Theorem}
\newtheorem*{cor*}{Corollary}

\definecolor{darkblue}{rgb}{0,0,0.7}

\tikzstyle{xshiftu}=[shift = {(#1, 0)}]
\tikzstyle{yshiftu}=[shift = {(0, #1)}]

\tikzstyle{dot}=[inner sep=0.25mm,minimum width=1mm,minimum height=1mm,draw,shape=circle,text depth=-0.2mm]
\tikzstyle{white dot}=[dot,fill=white, draw=black]
\tikzstyle{action}=[dot,fill=white,scale=0.667,inner sep=0.5mm]
\tikzstyle{copier}=[dot,fill=white,scale=2.0]
\tikzstyle{black copier}=[dot,fill=black,scale=2.0]

\tikzstyle{box}=[fill=white, draw=black, shape=rectangle]
\tikzstyle{medium box}=[fill=white, draw=black, shape=rectangle, minimum width=1.5cm, minimum height=0.66cm]
\tikzstyle{arrow box}=[fill=white, draw, shape=rectangle,minimum height=5mm,yshift=-0.5mm,minimum width=5mm]

\tikzstyle{effect}=[regular polygon, regular polygon sides=3,draw]
\tikzstyle{state0}=[regular polygon, regular polygon sides=3,draw,shape border rotate=0]
\tikzstyle{state90}=[regular polygon, regular polygon sides=3,draw,shape border rotate=90]
\tikzstyle{state180}=[regular polygon, regular polygon sides=3,draw,shape border rotate=180]
\tikzstyle{state270}=[regular polygon, regular polygon sides=3,draw,shape border rotate=270]

\tikzstyle{scalar}=[diamond,draw,inner sep=1pt]

\tikzstyle{discarder}=[my ground,draw,inner sep=0pt,minimum width=4.2pt,minimum height=11.2pt,anchor=input,rotate=90]
\tikzstyle{discarder0}=[my ground,draw,inner sep=0pt,minimum width=4.2pt,minimum height=11.2pt,anchor=input,rotate=0]

\tikzstyle{pointy1}=[->]
\tikzstyle{midpoint1}=[-, {postaction={decorate,decoration={markings, mark=at position .5 with {\arrow{>}}}}}]
\tikzstyle{midpointy1pointy}=[->, {postaction={decorate,decoration={markings, mark=at position .5 with {\arrow{>}}}}}]
\tikzstyle{dashed1}=[-, dashed]
\tikzstyle{dotted1}=[-, dotted]
\tikzstyle{dash-pointy}=[->, dashed]

\input{strings.tikzdefs}

\ifexternalizetikz\tikzexternaldisable\fi
\newsavebox\sbground
\savebox\sbground{%
  \begin{tikzpicture}[baseline=0pt]
    \draw (0,-.1ex) to (0,.85ex)
    node[ground IEC,draw,anchor=input,inner sep=0pt,
    minimum width=3.15pt,minimum height=8.4pt,rotate=90] {};
  \end{tikzpicture}%
}

\newsavebox\sbcopier
\savebox\sbcopier{%
  \begin{tikzpicture}[baseline=0pt]
    \node[copier,scale=0.7] (a) at (0,3.8pt) {};
    \draw (a) -- +(-90:.21);
    \draw (a) -- +(45:.21);
    \draw (a) -- +(135:.21);
  \end{tikzpicture}}

\ifexternalizetikz\tikzexternalenable\fi

\newsavebox\bsbcopier
\savebox\bsbcopier{%
  \begin{tikzpicture}[baseline=0pt]
    \node[black copier,scale=0.7] (a) at (0,3.8pt) {};
    \draw (a) -- +(-90:.21);
    \draw (a) -- +(45:.21);
    \draw (a) -- +(135:.21);
  \end{tikzpicture}}
\newcommand{\bcopier}{\mathord{\usebox\bsbcopier}}
\ifexternalizetikz\tikzexternalenable\fi

\usepackage{quiver}

\newcommand{\Poly}[1]{\Cat{Poly}_{\cat{#1}}}

\newcommand{\BLens}[1]{\Cat{BayesLens}_{\cat{#1}}} %
\newcommand{\BLCtx}[1]{\overline{\BLens{#1}}}    %

\newcommand{\SGame}[1]{\Cat{SGame}_{#1}}     %

\newcommand{\IntGame}[1]{\Cat{IntGame}_{#1}}     %
\newcommand{\SimpSGame}[1]{\Cat{SimpSGame}_{#1}} %

\newcommand{\PSGame}[1]{\Cat{PSGame}_{#1}}   %

\newcommand{\MProc}{\Cat{MrkProc}}
\newcommand{\MProcT}[1]{\MProc^{#1}}

\newcommand{\HiBi}{\Cat{HiBi}}
\newcommand{\HiBiD}{\HiBi(\Fun{D})}
\newcommand{\HiBiF}[1]{\HiBi\left({#1}\right)}

\newcommand{\deloop}[1]{\mathbf{B}#1}

\newcommand{\Sum}{\sum\limits}

\begin{document}
\maketitle

\begin{abstract}
  We extend our earlier work on the compositional structure of cybernetic
  systems in order to account for the embodiment of such systems. All their
  interactions proceed through their bodies' boundaries: sensations impinge on
  their surfaces, and actions correspond to changes in their configurations. We
  formalize this morphological perspective using polynomial functors. The
  `internal universes' of systems are shown to constitute an indexed category of
  statistical games over polynomials; their dynamics form an indexed category of
  behaviours. We characterize \emph{active inference doctrines} as indexed
  functors between such categories, resolving a number of open problems in our
  earlier work, and pointing to a formalization of the \emph{free energy
  principle} as adjoint to such doctrines. We illustrate our framework through
  fundamental examples from biology, including homeostasis, morphogenesis, and
  autopoiesis, and suggest a formal connection between spatial navigation and
  the process of proof.
\end{abstract}

\section{Introduction}

In a submission to ACT 2020 \cite{Smithe2020Cyber}, we presented some first
steps towards a theory of categorical cybernetics, motivated by concerns about
what gives physical systems life.  We explained that perception and action could
both be described as processes of Bayesian inference: on the one hand, adjusting
beliefs about the world on the basis of observational evidence; on the other,
adjusting the world itself in order better to match beliefs. In each case, the
system must instantiate a number of structures: a choice of `prior' belief about
the state of the world; a mechanism to generate predictions about sense-data on
the basis of that belief, called a `stochastic channel'; and a (typically
approximate) Bayesian inversion of that channel, by which to update those
beliefs, in light of sensory observations.

The pairing of a prior with a stochastic channel corresponds to what is called
in the informal literature a \emph{generative model}, and it is common to
suppose that these models are hierarchical: that is, that the stochastic channel
factors as some composite; one imagines that each factor corresponds to
predictions at some level of detail, cascading for example down from high-level
abstractions to individual photoreceptors. In our earlier submission, we
formalized this compositional structure using the bidirectional `lens'
pattern\footnote{See \cite{Smithe2020Bayesian} for a pedagogical presentation of
the fundamental structures.} --- since predictions and inversions are oppositely
directed --- and characterized a number of approximate inference processes using
a novel category of \emph{statistical games}, whose best responses correspond to
optimal inferences. A cybernetic system was then defined as a `dynamical
realisation' of such a game.

This formalism left some things to be desired: our notion of dynamical
realisation was ill-defined, and the notion of `action' was overly abstract. In
this submission, we resolve these issues, substantially simplifying our
presentation along the way. We explain that various recipes for performing
approximate inference, corresponding to our earlier informal notion of dynamical
realisation, form functorial \emph{approximate inference doctrines}, between
appropriate categories of statistical games and dynamical systems. Then, to
formalize a satisfactory notion of action, we note that any active system has a
boundary defining its morphology, and that it acts by changing the shape of this
boundary; in order to \emph{act on} another system, it couples part of this
boundary to that other system, so as to change the composite shape.

To formalize the shapes of systems and their interactions, we adopt polynomial
functors: each polynomial will encode the `phenotype' (possible shapes or
configurations) of a system, and the sensorium possible in each
configuration. To give such systems life, we construct categories of statistical
games and dynamical behaviours indexed by polynomials. An \emph{active inference
doctrine} is then an indexed functor between such categories. This framework
enables a number of possibilities: we can construct a generative model for a
corporation on the basis of models for its employees; we give compositional
descriptions of fundamental processes of life such as homeostasis and
morphogenesis, and point towards an account of autopoiesis; and we sketch the
process by which living systems internalize the structures of their
environments, and navigate accordingly, noting that such navigation in abstract
spaces corresponds precisely to the process of proof.

The work presented here is work in progress, and owing to constraints of space
and time, it has not been possible to elaborate everything that we would have
liked; this means that we defer proofs of the main results to subsequent
elaborations of this extended abstract. Two such elaborations are our series of
papers on compositional active inference, beginning with
\cite{Smithe2021Compositional1}, and our work on open dynamical systems with
polynomial interfaces \cite{Smithe2022Open}. Notwithstanding these limitations,
we believe the results here go some of the way to answering the open questions,
of elegance and interaction, sketched at the end of our earlier submission. We
see this work as making baby steps towards a theory of \emph{embodied}
cybernetics.

\paragraph{Acknowledgements}

We thank the members of the Topos Institute for stimulating discussions, and the
Foundational Questions Institute and Topos Institute for financial support.

\section{Simpler Statistical Games}

We begin by sketching a refinement of the statistical games formalism developed
in \cite{Smithe2020Cyber}.  For much more detail, we refer the reader to
\cite{Smithe2021Compositional1}.  First, we recall the bidirectional structure
of Bayesian inversion.

\subsection{Bayesian Lenses}

\begin{defn}[{\cite[Def.~3.3]{Spivak2019Generalized}}]
  The category \(\Cat{GrLens}_F\) of \textit{Grothendieck lenses} for a
  pseudofunctor \(F : \cat{C}\op \to \Cat{Cat}\) is the total category of the
  Grothendieck construction for the pointwise opposite of \(F\).
\end{defn}

\begin{prop}[\(\Cat{GrLens}_F\) is a category] \label{prop:grlens-cat}
  The objects \((\Cat{GrLens}_F)_0\) of \(\Cat{GrLens}_F\) are (dependent) pairs
  \((C, X)\) with \(C : \cat{C}\) and \(X : F(C)\), and its hom-sets
  \(\Cat{GrLens}_F \big( (C, X), (C', X') \big)\) are dependent sums
  \begin{equation*}
    \Cat{GrLens}_F \big( (C, X), (C', X') \big)
    = \sum_{f \, : \, \cat{C}(C, C')} F(C) \big( F(f)(X'), X \big)
  \end{equation*}
  so that a morphism \((C, X) \lensto (C', X')\) is a pair \((f, f^\dag)\) of
  \(f : \cat{C}(C, C')\) and \(f^\dag : F(C) \big( F(f)(X'), X \big)\). We call
  such pairs \textit{Grothendieck lenses for} \(F\) or \(F\)\textit{-lenses}.
  \begin{proof}[Proof sketch]
    The identity Grothendieck lens on \((C, X)\) is \(\id_{(C, X)} = (\id_C,
    \id_X)\). Sequential composition is as follows. Given \((f, f^\dag) : (C, X)
    \lensto (C', X')\) and \((g, g^\dag) : (C', X') \lensto (D, Y)\), their
    composite \((g, g^\dag) \lenscirc (f, f^\dag)\) is defined to be the lens
    \(\big(g \circ f, f^\dag \circ F(f)(g^\dag) \big) : (C, X) \lensto (D,
    Y)\). Associativity and unitality of composition follow from the
    functoriality of \(F\).
  \end{proof}
\end{prop}

\begin{defn} \label{def:simp-lens}
  Suppose \(F(C)_0 = \cat{C}_0\), with \(F : \cat{C}\op \to \Cat{Cat}\) a
  pseudofunctor. Define \(\Cat{SimpGrLens}_F\) to be the full subcategory of
  \(\Cat{GrLens}_F\) whose objects are duplicate pairs \((C, C)\) of objects
  \(C\) in \(\cat{C}\). We call \(\Cat{SimpGrLens}_F\) the category of
  \textit{simple} \(F\)-lenses. More generally, any lens between such duplicate
  pairs will be called a \textit{simple lens}. Since duplicating the objects in
  the pairs \((C,C)\) is redundant, we will write the objects simply as \(C\).
\end{defn}

\begin{defn} \label{def:stat-cat}
  Let \((\cat{C}, \otimes, I)\) be a monoidal category enriched in a Cartesian
  closed category \(\Cat{V}\). Define the \(\cat{C}\)\textit{-state-indexed}
  category \(\Fun{Stat}: \cat{C}\op \to \Cat{V\mdash Cat}\) as follows.  {\small
  \begin{align}
    \Fun{Stat} \;\; : \;\; \cat{C}\op \; & \to \; \Cat{V\mdash Cat} \nonumber \\
    X & \mapsto \Fun{Stat}(X) := \quad \begin{pmatrix*}[l]
      & \Fun{Stat}(X)_0 & := \quad \;\;\; \cat{C}_0 \\
      & \Fun{Stat}(X)(A, B) & := \quad \;\;\; \Cat{V}(\cat{C}(I, X), \cat{C}(A, B)) \\
      \id_A \: : & \Fun{Stat}(X)(A, A) & := \quad
      \left\{ \begin{aligned}
        \id_A : & \; \cat{C}(I, X)     \to     \cat{C}(A, A) \\
        & \quad\;\;\: \rho \quad \mapsto \quad \id_A
      \end{aligned} \right. \label{eq:stat} \\
    \end{pmatrix*} \\ \nonumber \\
    f : \cat{C}(Y, X) & \mapsto \begin{pmatrix*}[c]
      \Fun{Stat}(f) \; : & \Fun{Stat}(X) & \to & \Fun{Stat}(Y) \vspace*{0.5em} \\
      & \Fun{Stat}(X)_0 & = & \Fun{Stat}(Y)_0 \vspace*{0.5em} \\
      & \Cat{V}(\cat{C}(I, X), \cat{C}(A, B)) & \to & \Cat{V}(\cat{C}(I, Y), \cat{C}(A, B)) \vspace*{0.125em} \\
      & \alpha & \mapsto & f^\ast \alpha : \big( \, \sigma : \cat{C}(I, Y) \, \big) \mapsto \big( \, \alpha(f \klcirc \sigma) : \cat{C}(A, B) \, \big)
    \end{pmatrix*} \nonumber
  \end{align}
  }%
  Composition in each fibre \(\Fun{Stat}(X)\) is as in \(\cat{C}\). Explicitly,
  indicating morphisms \(\cat{C}(I, X) \to \cat{C}(A, B)\) in \(\Fun{Stat}(X)\)
  by \(A \xklto{X} B\), and given \(\alpha : A \xklto{X} B\) and \(\beta : B
  \xklto{X} C\), their composite is \(\beta \circ \alpha : A \xklto{X} C := \rho
  \mapsto \beta(\rho) \klcirc \alpha(\rho)\), where here we indicate composition
  in \(\cat{C}\) by \(\klcirc\) and composition in the fibres \(\Fun{Stat}(X)\)
  by \(\circ\). Given \(f : Y \klto X\) in \(\cat{C}\), the induced functor
  \(\Fun{Stat}(f) : \Fun{Stat}(X) \to \Fun{Stat}(Y)\) acts by pullback.
\end{defn}

\begin{ex}
  Typically, a choice of \(\cat{C}\) above will be the Kleisli category
  \(\Kl(\Pa)\) of a probability monad, such as the Giry monad \(\Giry :
  \Cat{Meas}\to\Cat{Meas}\) on measurable spaces; and \(\Cat{V}\) will either be
  \(\Set\) or (better) some `nice' category of measurable spaces. However, the
  category \(\Cat{Meas}\) of general measurable spaces is not Cartesian closed,
  as there is no general way to make the evaluation maps \(\Cat{Meas}(X, Y)
  \times X \to Y\) measurable, meaning that if we take \(\cat{C} = \Kl(\Giry)\)
  above, then we are forced to take \(\Cat{V} = \Set\). In turn, this makes the
  inversion maps \(c^\dag : \Kl(\Giry)(1, X) \to \Kl(\Giry)(Y,X)\) into mere
  functions. We can salvage measurability by working instead with \(\Kl(\Qa)\),
  where \(\Qa : \Cat{QBS} \to \Cat{QBS}\) is the analogue of the Giry monad for
  quasi-Borel spaces \cite{Heunen2017Convenient}. The category \(\Cat{QBS}\) is
  indeed Cartesian closed, and \(\Kl(\Qa)\) is enriched in \(\Cat{QBS}\), so
  that we can instantiate \(\Fun{Stat}\) there, and the corresponding inversion
  maps are accordingly measurable.
\end{ex}

We define the category of Bayesian lenses in \(\cat{C}\) to be the category of
\(\Fun{Stat}\)-lenses.

\begin{defn} \label{def:stat-lens}
  The category \(\BLens{C}\) of Bayesian lenses in \(\cat{C}\) is the category
  \(\Cat{GrLens}_{\Fun{Stat}}\) of Grothendieck lenses for the functor
  \(\Fun{Stat}\). A \textit{Bayesian lens} is a morphism in \(\BLens{C}\). Where
  the category \(\cat{C}\) is evident from the context, we will just write
  \(\Cat{BayesLens}\).
\end{defn}

Unpacking this definition, we find that the objects of \(\BLens{C}\) are pairs
\((X, A)\) of objects of \(\cat{C}\). Morphisms (that is, Bayesian lenses) \((X,
A) \lensto (Y, B)\) are pairs \((c, c^\dag)\) of a channel \(c : X \klto Y\) and
a ``generalized Bayesian inversion'' \(c^\dag : B \xklto{X} A\); that is,
elements of the hom objects
\begin{align*}
  \BLens{C}\big((X,A),(Y,B)\big)
  :&= \Cat{GrLens}_\Fun{Stat} \big((X,A),(Y,B)\big) \\
  &\cong \cat{C}(X, Y) \times \Cat{V} \big( \cat{C}(I, X), \cat{C}(B, A) \big) \, .
\end{align*}
The identity Bayesian lens on \((X, A)\) is \((\id_X, \id_A)\), where by abuse
of notation \(\id_A : \cat{C}(I, Y) \to \cat{C}(A, A)\) is the constant map
\(\id_A\) defined in equation \eqref{eq:stat} that takes any state on \(Y\) to
the identity on \(A\).

The sequential composite \((d, d^\dag) \lenscirc (c, c^\dag)\) of \((c, c^\dag)
: (X, A) \lensto (Y, B)\) and \((d, d^\dag) : (Y, B) \lensto (Z, C)\) is the
Bayesian lens \(\big( (d \klcirc c), (c^\dag \circ c^\ast d^\dag) \big) : (X, A)
\lensto (Z, C)\) where \((c^\dag \circ c^\ast d^\dag) : C \xklto{X} A\) takes a
state \(\pi : I \klto X\) to the channel \(c^\dag_{\pi} \klcirc \d^\dag_{c
  \klcirc \pi} : C \klto A\).

\begin{defn}
  Given a Bayesian lens \((c,c') : (X,A)\lensto (Y,B)\), we will call \(c\) its
  \textit{forwards} or \textit{prediction} channel and \(c'\) its
  \textit{backwards} or \textit{update} channel (even though \(c'\) is really a
  family of channels).
\end{defn}

\begin{defn}[Bayesian inversion] \label{def:admit-bayes}
  We say that a channel \(c : X \klto Y\) \textit{admits Bayesian inversion}
  with respect to \(\pi : I \klto X\) if there exists a channel \(c^\dag_\pi : Y
  \klto X\), called the \textit{Bayesian inversion of} \(c\) with respect to
  \(\pi\), satisfying the following equation \cite[eq. 5]{Cho2017Disintegration}
  in the graphical calculus of \(\cat{C}\):
  \begin{equation} \label{eq:bayes-abstr}
    \scalebox{0.75}{\begin{tikzpicture}
	\begin{pgfonlayer}{nodelayer}
		\node [style=black copier] (1) at (0, 0) {};
		\node [style=none] (2) at (-2, 6) {};
		\node [style=arrow box] (3) at (2, 3.5) {$c$};
		\node [style=none] (4) at (2, 6) {};
		\node [style=none] (5) at (-2, 2) {};
		\node [style=state180] (6) at (0, -4.5) {$\pi$};
		\node [style=none] (7) at (-1.5, 5.75) {$X$};
		\node [style=none] (8) at (2.5, 5.75) {$Y$};
		\node [style=none] (9) at (2, 2) {};
	\end{pgfonlayer}
	\begin{pgfonlayer}{edgelayer}
		\draw [bend left=45] (1) to (5.center);
		\draw (5.center) to (2.center);
		\draw (3) to (4.center);
		\draw (6) to (1);
		\draw [bend right=45] (1) to (9.center);
		\draw (9.center) to (3);
	\end{pgfonlayer}
\end{tikzpicture}
}
    \quad = \quad
    \scalebox{0.75}{\begin{tikzpicture}
	\begin{pgfonlayer}{nodelayer}
		\node [style=black copier] (0) at (0, 0) {};
		\node [style=none] (1) at (2, 6) {};
		\node [style=arrow box] (2) at (-2, 3.5) {$c^\dag_\pi$};
		\node [style=none] (3) at (-2, 6) {};
		\node [style=none] (4) at (2, 2) {};
		\node [style=state180] (5) at (0, -4.5) {$\pi$};
		\node [style=arrow box] (6) at (0, -2) {$c$};
		\node [style=none] (7) at (-1.5, 5.75) {$X$};
		\node [style=none] (8) at (2.5, 5.75) {$Y$};
		\node [style=none] (9) at (-2, 2) {};
	\end{pgfonlayer}
	\begin{pgfonlayer}{edgelayer}
		\draw [bend right=45] (0) to (4.center);
		\draw (4.center) to (1.center);
		\draw (2) to (3.center);
		\draw (5) to (6);
		\draw (6) to (0);
		\draw [bend left=45] (0) to (9.center);
		\draw (9.center) to (2);
	\end{pgfonlayer}
\end{tikzpicture}
}
  \end{equation}
  We say that \(c\) admits Bayesian inversion \textit{tout court} if \(c\)
  admits Bayesian inversion with respect to all states \(\pi : I \klto X\) such
  that \(c \klcirc \pi\) has non-empty support. We say that a category
  \(\cat{C}\) admits Bayesian inversion if all its morphisms admit Bayesian
  inversion \textit{tout court}.
\end{defn}

\begin{defn} \label{def:gen-model}
  We call the pairing \((\pi, c)\) of a state \(\pi : I \klto X\) with a channel
  \(c : X \klto Y\) a \textit{generative model} \(X \klto Y\). It induces a
  joint distribution \(\omega_{(\pi, c)} := (\id_X\otimes\, c) \klcirc
  \bcopier_X \klcirc \pi : I \klto X \otimes Y\).
\end{defn}

\begin{defn}
  Let \((c, c^\dag) : (X, X) \lensto (Y, Y)\) be a Bayesian lens. We say that
  \((c, c^\dag)\) is \textit{exact} if \(c\) admits Bayesian inversion and, for
  each \(\pi : I \klto X\) such that \(c \klcirc \pi\) has non-empty support,
  \(c\) and \(c^\dag_\pi\) together satisfy equation \eqref{eq:bayes-abstr}.
  Bayesian lenses that are not exact are said to be \textit{approximate}.
\end{defn}

\begin{defn}[Almost-equality]
  Given a state \(\pi : I \klto X\), we say that two parallel channels \(c,d : X
  \klto Y\) are \(\pi\)\textit{-almost-equal}, denoted \(c \overset{\pi}{\sim}
  d\), if the joint distributions of the two generative models \((\pi, c)\) and
  \((\pi, d)\) are equal; that is, if \((\id\otimes\, c) \klcirc \bcopier
  \klcirc \pi = (\id\otimes\, d) \klcirc \bcopier \klcirc \pi\).
\end{defn}

\begin{thm}[\cite{Smithe2020Bayesian}] \label{thm:buco}
  Let \((c, c^\dag)\) and \((d, d^\dag)\) be sequentially composable exact
  Bayesian lenses. Then the contravariant component of the composite lens \((d,
  d^\dag) \lenscirc (c, c^\dag) = (d \klcirc c, c^\dag \circ c^\ast d^\dag)\)
  is, up to \(d \klcirc c \klcirc \pi\)-almost-equality, the Bayesian inversion
  of \(d \klcirc c\) with respect to any state \(\pi\) on the domain of \(c\)
  such that \(c \klcirc \pi\) has non-empty support.
\end{thm}

\subsection{Statistical Games}

The performance of a statistical or cybernetic system depends upon its
interaction with its environment, and the prior beliefs that it started with. We
will therefore define a statistical game to be a Bayesian lens paired with a
fitness function measuring performance in context; and for this we need a notion
of context. For this notion, we take inspiration from compositional game theory
\cite{Bolt2019Bayesian}.

\begin{defn}
  A \textit{context} for a Bayesian lens is an element of the profunctor
  \(\BLCtx{C}\) defined by
  \[
  \begin{matrix*}[c]
    \BLens{C} & \times & \BLens{C}\op &   \to   & \Cat{V} \\
        {-}   & \times &      {=}     & \mapsto & \BLens{C}((I,I),{-})\times\BLens{C}({=},(I,I)) \\
  \end{matrix*}
  \]
  where \(I\) is the monoidal unit in \(\cat{C}\), \(\BLens{C}((I,I),{-}) :
  \BLens{C} \to \Cat{V}\) is the representable \mbox{copresheaf} on \((I,I)\),
  and \(\BLens{C}({=},(I,I)) : \BLens{C}\op \to \Cat{V}\) is the representable
  presheaf. Hence a context for a lens \((X,A)\lensto(Y,B)\) is an element of
  \[
  \BLCtx{C}\big((X,A),(Y,B)\big) := \BLens{C}\big((I,I),(X,A))\times\BLens{C}((Y,B),(I,I)\big) \, .
  \]
\end{defn}

\begin{prop} \label{prop:ctx-terminal}
  When \(I\) is terminal in \(\cat{C}\) and the base \(\Cat{V}\) of enrichment
  of \(\cat{C}\) is well pointed (as when \(\Cat{V} = \Set\)), we have
  \(\BLCtx{C}\big((X,A),(Y,B)\big) \cong \cat{C}(I,X)\times\Cat{V}\bigl(\cat{C}(I,B),\cat{C}(I,Y)\bigr)\).
  \begin{proof}
    A straightforward calculation which we omit: use that \(X\xto{c}Y\xto{!}I =
    X\xto{!}I\) for any \(c : X\to Y\).
  \end{proof}
\end{prop}

\begin{prop} \label{prop:local-ctx}
  Given a context \((\pi, k) : \BLCtx{C}\big((X,A),(Z,C)\big)\) for a composite
  lens \((X,A) \overset{f}{\lensto} (Y,B) \overset{g}{\lensto} (Z,C)\), we
  obtain contexts for the factors:
  \begin{align*}
    \BLCtx{C}\big((X,A),g\big)(\pi,k) = (\pi, k\lenscirc g) : \BLCtx{C}\big((X,A),(Y,B)\big) \, , \\
    \BLCtx{C}\big(f,(Z,C)\big)(\pi,k) = (f \lenscirc \pi, k) : \BLCtx{C}\big((Y,B),(Z,C)\big) \, .
  \end{align*}
\end{prop}

We are now in a position to define the category of statistical games over
$\cat{C}$.

\begin{prop} \label{prop:cat-sgame}
  Let \(\cat{C}\) be a \(\Cat{V}\)-category admitting Bayesian inversion and let
  \((R, +, 0)\) be a monoid in \(\Cat{V}\). Then there is a category
  \(\SGame[R]{\cat{C}}\) whose objects are the objects of \(\BLens{C}\) and
  whose morphisms \((X, A) \to (Y, B)\) are \textit{statistical games}: pairs
  \((f, \phi)\) of a lens \(f : \BLens{C}\big((X, A), (Y, B)\big)\) and a
  \textit{fitness function} \(\phi : \BLCtx{C}\big((X,A),(Y,B)\big) \to R\).
  When \(R\) is the monoid of reals \(\rr\), then we just denote the category by
  \(\SGame{\cat{C}}\).
  \begin{proof}
    Suppose given statistical games \((f,\phi) : (X,A) \to (Y,B)\) and
    \((g,\psi) : (Y,B) \to (Z,C)\). We seek a composite game \((g,\psi) \circ
    (f,\phi) := (gf,\psi\phi) : (X,A) \to (Z,C)\). We have \(gf = g \lenscirc
    f\) by lens composition. Propositon \ref{prop:local-ctx} gives us a family
    of functions \(\mathsf{localCtx}\) with signature
    \begin{align*}
    & \BLCtx{C}\big((X, A), (Z, C)\big)
    \times \BLens{C}\big((X, A), (Y, B)\big)
    \times \BLens{C}\big((Y, B), (Z, C)\big) \\
    & \to \BLCtx{C}\big((X, A), (Y, B)\big) \times
          \BLCtx{C}\big((Y, B), (Z, C)\big) \, .
    \end{align*}
    We therefore identify the composite fitness function \(\psi\phi\) as
    \[
    \psi\phi := + \circ (\phi, \psi) \circ \mathsf{localCtx}(-, f, g)
    \]
    where \(+ : R \times R \to R\) is the monoid operation. The identity game
    \((X,A) \to (X,A)\) is given by \((\id, 0)\), the pairing of the identity
    lens on \((X, A)\) with the unit \(0\) of the monoid \(R\). Associativity
    and unitality are immediate from lens composition and the monoid laws.
  \end{proof}
\end{prop}

\begin{defn} \label{def:simp-sgame}
  We will write \(\SimpSGame{\cat{C}} \hookrightarrow \SGame{\cat{C}}\) for the
  full subcategory of \(\SGame{\cat{C}}\) defined on simple Bayesian lenses
  \((X,X) \lensto (Y, Y)\). As in the case of simple lenses (Definition
  \ref{def:simp-lens}), we will eschew redundancy by writing the objects
  \((X,X)\) simply as \(X\).
\end{defn}

We now present some key examples of statistical games.

\begin{ex}[Bayesian inference] \label{ex:bayes-game}
  Let \(D : \cat{C}(I,X) \times \cat{C}(I,X) \to \rr\) be a measure of
  divergence between states on \(X\). Then a (simple) \(D\)\textit{-Bayesian
    inference} game is a statistical game \((X,X) \to (Y,Y)\) with fitness
  function \(\phi : \BLCtx{C}\bigl((X,X),(Y,Y)\bigr) \to \rr\) given by
  \(\phi(\pi,k) = \E_{y\sim k\klcirc c\klcirc\pi} \left[ D\left(c'_\pi(y),
    c^\dag_\pi(y)\right) \right]\), where \((c,c')\) constitutes the lens part
  of the game and \(c^\dag_\pi\) is the exact inversion of \(c\) with respect to
  \(\pi\).
\end{ex}

Note that we say that \(D\) is a ``measure of divergence between states on
\(X\)''. By this we mean any function of the given type with the semantical
interpretation that it acts like a distance measure between states. But this is
not to say that \(D\) is a metric or even pseudometric. One usually requires
that \(D(\pi, \pi') = 0 \iff \pi = \pi'\), but typical choices do not also
satisfy symmetry nor subadditivity. An important such typical choice is the
\textit{relative entropy} or \textit{Kullback-Leibler divergence}, denoted
\(D_{KL}\).

\begin{defn} \label{def:d-kl}
  The \textit{Kullback-Leibler divergence} \(D_{KL} : \cat{C}(I, X) \times
  \cat{C}(I, X) \to \rr\) is defined by
  \[
  D_{KL}(\alpha, \beta) := \E_{x \sim \alpha}[\log p (x)] - \E_{x \sim \alpha}[\log q (x)]
  \]
  where \(p\) and \(q\) are density functions corresponding to the states
  \(\alpha\) and \(\beta\).
\end{defn}

Typically, computing $D_{KL}(c'_\pi(x), c^\dag_\pi(x))$ is computationally
difficult, so one resorts to optimizing an upper bound; a prominent choice is
the \emph{free energy} \cite{Friston2007Variational}.

\begin{defn}[\(D\)-free energy] \label{def:d-free-energy}
  Let \((\pi, c)\) be a generative model with \(c : X \klto Y\). Let \(p_c : Y
  \times X \to \rr_+\) and \(p_\pi : X \to \rr_+\) be density functions
  corresponding to \(c\) and \(\pi\). Let \(p_{c \klcirc \pi} : Y \to \rr_+\) be
  a density function for the composite \(c\klcirc\pi\).  Let \(c'_\pi\) be a
  channel \(Y \klto X\) that we take to be an approximation of the Bayesian
  inversion of \(c\) with respect to \(\pi\) and that admits a density function
  \(q : X \times Y \to \rr_+\). Finally, let \(D : \cat{C}(I, X) \times
  \cat{C}(I, X) \to \rr\) be a measure of divergence between states on \(X\).
  Then the \(D\)\textit{-free energy} of \(c'_\pi\) with respect to the
  generative model given an observation \(y:Y\) is the quantity
  \begin{equation} \label{eq:free-energy}
  \Fa_D(c'_\pi, c, \pi, y) := \E_{x \sim c'_\pi(y)} \left[ - \log p_c(y | x)  \right] + D\left(c'_\pi(y), \pi\right) \, .
  \end{equation}
  We will elide the dependence on the model when it is clear from the context,
  writing only \(\Fa_D(y)\).
\end{defn}

The \(D\)-free energy is an upper bound on \(D\) when \(D\) is the relative
entropy \(D_{KL}\).

\begin{prop}[Evidence upper bound] \label{prop:eubo}
  The \(D_{KL}\)-free energy satisfies the following equality:
  \[
  \Fa_{D_{KL}}(y) = D_{KL}\left[ c'_\pi(y), c^\dag_\pi(y)\right] - \log p_{c \klcirc \pi}(y) = \E_{x \sim c'_\pi(y)} \left[ \log \frac{q(x|y)}{p_c(y|x) \cdot p_\pi(x)} \right]
  \]
  Since \(\log p_{c \klcirc \pi}(y)\) is always negative, the free energy is an
  upper bound on \(D_{KL}\left[ c'_\pi(y), c^\dag_\pi(y)\right]\), where
  \(c^\dag_\pi\) is the exact Bayesian inversion of the channel \(c\) with
  respect to the prior \(\pi\). Similarly, the free energy is an upper bound on
  the negative log-likelihood \(-\log p_{c \klcirc \pi}(y)\). Thinking of this
  latter quantity as a measure of the ``model evidence'' gives us the
  alternative name \textit{evidence upper bound} for the \(D_{KL}\)-free energy.
\end{prop}

Any system that performs (approximate) Bayesian inversion can thus be seen as
minimizing some free energy. The \emph{free energy principle} says that all it
means to be an adaptive system is to embody a process of approximate inference
in this way. We therefore define free energy games:

\begin{ex}[Free energy game] \label{ex:simp-autoenc-game}
  Let \(D : \cat{C}(I, X) \times \cat{C}(I, X) \to \rr\) be a measure of
  divergence between states on \(X\). Then a simple \(D\)\textit{-free energy
    game} is a simple statistical game \((X, X) \to (Y, Y)\) with fitness
  function \(\phi : \BLCtx{C}\big((X, X), (Y, Y)\big) \to \rr\) given by
  \(\phi(\pi, k) = \E_{y\sim k\klcirc c\klcirc\pi} \left[ \Fa_{D} \left(c'_\pi,
    c, \pi, y\right) \right]\)
  where \((c, c') : (X,X) \lensto (Y,Y)\) constitutes the lens part of the game.
\end{ex}

\begin{rmk} \label{rmk:para}
  It is also often of interest to consider \emph{parameterized} channels, for
  which we can use the $\Cat{Para}$ construction \cite{Spivak2021Learners}. This
  acts by adjoining an object of parameters to the domain of the category at
  hand (such as the category of Bayesian lenses), and tensoring parameters of
  composite morphisms. Formally, this corresponds to a generalization of the
  indexed category of state-dependent morphisms, and forms the subject of
  another paper in these proceedings \cite{Capucci2021Towards}. The parameters
  might represent the `weights' of a neural network, or encode some structure
  about the possible predictions. Lacking the space to do justice to this
  structure here, we nonetheless leave it at that, and refer the reader to our
  paper \cite{Smithe2021Compositional1} for more details.
\end{rmk}

\section{Systems Within Interfaces; Worlds Within Worlds}

In this section, we develop the structures required for extending the formalism
of statistical games to embodied systems.

\subsection{Polynomials for Embodiment and Interaction}

Each system in our universe inhabits some interface or boundary. It receives
signals from its environments through this boundary, and can act by changing its
shape (and, as we will see later, its position). As a system changes its shape,
the set of possible immanent signals might change accordingly: consider a
hedgehog rolling itself into a ball, thereby protecting its soft underbelly from
harm (amongst other immanent signals). A system may also change its shape by
coupling itself to some other system, such as when we pick up chalk to work
through a problem. And shapes can be abstract: we change our `shapes' when we
enter an online video conference, or move within a virtual reality. We describe
all of these interactions formally using polynomial functors, drawing on the
work of \cite{Spivak2020Poly}.

\begin{defn}
  Let $\cat{E}$ be a locally Cartesian closed category, and denote by $y^A$ the
  representable copresheaf $y^A := \cat{E}(A, -) : \cat{E} \to \cat{E}$. A
  \emph{polynomial functor} $p$ is a coproduct of representable functors,
  written $p := \sum_{i : p(1)} y^{p_i}$, where $p(1) : \cat{E}$ is the indexing
  object. The category of polynomial functors in $\cat{E}$ is the full
  subcategory $\Poly{E} \hookrightarrow [\cat{E}, \cat{E}]$ of the
  $\cat{E}$-copresheaf category spanned by coproducts of representables. A
  morphism of polynomials is therefore a natural transformation.
\end{defn}

\begin{rmk} %
  Every polynomial functor $P : \cat{E} \to \cat{E}$ corresponds to a bundle $p
  : E \to B$ in $\cat{E}$, for which $B = P(1)$ and for each $i : P(1)$, the
  fibre $p_i$ is $P(i)$. We will henceforth elide the distinction between a
  copresheaf $P$ and its corresponding bundle $p$, writing $p(1) := B$ and $p[i]
  := p_i$, where $E = \sum_i p[i]$. A natural transformation $f : p \to q$
  between copresheaves therefore corresponds to a map of bundles. In the case of
  polynomials, by the Yoneda lemma, this map is given by a `forwards' map $f_1 :
  p(1) \to q(1)$ and a family of `backwards' maps $f^\# : q[f_1(\mdash)] \to
  p[\mdash]$ indexed by $p(1)$, as in the left diagram below. Given $f : p \to
  q$ and $g : q \to r$, their composite $g \circ f : p \to r$ is as in the right
  diagram below.
  \begin{equation*}
    \begin{tikzcd}
      E & {f^*F} & F \\
      B & B & C
      \arrow["{f^\#}"', from=1-2, to=1-1]
      \arrow[from=1-2, to=1-3]
      \arrow["q", from=1-3, to=2-3]
      \arrow["p"', from=1-1, to=2-1]
      \arrow[from=2-1, to=2-2, Rightarrow, no head]
      \arrow["{f_1}", from=2-2, to=2-3]
      \arrow[from=1-2, to=2-2]
      \arrow["\lrcorner"{anchor=center, pos=0.125}, draw=none, from=1-2, to=2-3]
    \end{tikzcd}
    \qquad\qquad
    \begin{tikzcd}
      E & {f^*g^*G} & G \\
      B & B & D
      \arrow["{(gf)^\#}"', from=1-2, to=1-1]
      \arrow[from=1-2, to=1-3]
      \arrow["r", from=1-3, to=2-3]
      \arrow["p"', from=1-1, to=2-1]
      \arrow[from=2-2, to=2-1, Rightarrow, no head]
      \arrow["{g_1 \circ f_1}", from=2-2, to=2-3]
      \arrow[from=1-2, to=2-2]
      \arrow["\lrcorner"{anchor=center, pos=0.125}, draw=none, from=1-2, to=2-3]
    \end{tikzcd}
  \end{equation*}
  Here, $(gf)^\#$ is given by the $p(1)$-indexed family of composite maps
  $r[g_1(f_1(\mdash))] \xto{f^\ast g^\#} q[f_1(\mdash)] \xto{f^\#} p[\mdash]$.
\end{rmk}

In our morphological semantics, we will call a polynomial $p$ a
\emph{phenotype}, its base type $p(1)$ its \emph{morphology} and the total space
$\sum_i p[i]$ its \emph{sensorium}. We will call elements of the morphology
\emph{shapes} or \emph{configurations}, and elements of the sensorium
\emph{immanent signals}.

\begin{prop}[\cite{Spivak2020Poly}]
  There is a monoidal structure $(\Poly{E}, \otimes, y)$ that we interpret as
  ``putting systems in parallel''. Given $p : \sum_i p[i] \to p(1)$ and $q :
  \sum_j q[j] \to q(1)$, we have $p \otimes q = \sum_i \sum_j p[i] \times q[j]
  \to p(1) \times q(1)$. $y : 1 \to 1$ is then clearly unital. \qed
\end{prop}

\begin{prop}[\cite{Spivak2020Poly}]
  The monoidal structure $(\Poly{E}, \otimes, y)$ is closed, with corresponding
  internal hom denoted $[-, -]$.
\end{prop}

We interpret morphisms $(f_1, f^\#)$ of polynomials as encoding interaction
patterns; in particular, such morphisms encode how composite systems act as
unities. For example, a morphism $f : p \otimes q \to r$ specifies how the
systems $p$ and $q$ come together to form a system $r$: the map $f_1$ encodes
how $r$-configurations are constructed from configurations of $p$ and $q$; and
the map $f^\#$ encodes how immanent signals on $p$ and $q$ result from signals
on $r$ or from the interaction of $p$ and $q$. For intuition, consider two
people engaging in a handshake, or an enzyme acting on a protein to form a
complex. The internal hom $[o, p]$ encodes all the possible ways that an
$o$-phenotype system can ``plug into'' a $p$-phenotype system.

\begin{rmk}
  In the literature on active inference and the free energy principle, there is
  much debate about the concept and role of `Markov blankets', which are
  sometimes conceived to represent the boundary of an adaptive system. We
  believe that the algebra of polynomials will help to make such thinking
  precise, where it departs from the established usage of Markov blankets in
  Bayesian networks.
\end{rmk}

\subsection{Dynamical Systems on Polynomial Interfaces}

Although some dynamical systems can be modelled within $\Poly{E}$ itself, we
prefer to take a fibrational perspective, following the idea that polynomials
represent the boundaries of systems, separating `internal' states from
`external'. We will therefore adopt a pattern of indexing categories by
polynomials: in the case of dynamics, the fibre over a polynomial will be a
category of possible internal systems, whose projection forgets the internal
structure. We can only sketch the relevant structures here, and so refer the
reader to our subsequent paper \cite{Smithe2022Open} where we give a general
account of open dynamical systems as coalgebras for polynomial functors.

\begin{defn} \label{def:mproc}
  Let \(\Pa : \cat{E} \to \cat{E}\) be a probability monad on the category
  \(\cat{E}\), and let \(p : \Poly{E}\) be a polynomial in \(\cat{E}\). Let
  \((\Tt, +, 0)\) be a monoid in \(\cat{E}\), representing time. Then an
  \textbf{open Markov process} on the interface \(p\) with time \(\Tt\) consists
  in a triple \(\vartheta := (S, \vartheta^o, \vartheta^u)\) of a \textbf{state
    space} \(S : \cat{E}\) and two morphisms \(\vartheta^o : \Tt \times S \to
  p(1)\) and \(\vartheta^u : \sum_{t:\Tt} \sum_{s:\mathbb{S}}
  p[\vartheta^o(t,s)] \to \Pa S\), such that for any section \(\sigma : p(1)
  \to \sum_{i:p(1)} p[i]\) of \(p\), the maps \(\vartheta^\sigma : \Tt \times S
  \to \Pa S\) given by
  \[
  \Sum_{t:\Tt} S \xto{\vartheta^o(-)^\ast \sigma} \Sum_{t:\Tt} \Sum_{s:S} p[\vartheta^o(-, s)] \xto{\vartheta^u} \Pa S
  \]
  constitute an object in the functor category \(\Cat{Cat}\big(\deloop{\Tt},
  \Kl(\Pa)\big)\), where \(\deloop{\Tt}\) is the delooping of \(\Tt\) and
  \(\Kl(\Pa)\) is the Kleisli category of \(\Pa\). We call \(\vartheta^\sigma\)
  the \textbf{closure} of \(\vartheta\) by \(\sigma\).
\end{defn}

\begin{prop} \label{prop:mproc-cat}
  Open Markov processes over \(p\) with time \(\Tt\) form a category, denoted
  \(\MProcT{\Tt}_{\Pa}(p)\). Its morphisms are defined as follows.  Let
  \(\vartheta := (X, \vartheta^o, \vartheta^u)\) and \(\psi := (Y, \psi^o,
  \psi^u)\) be two Markov processes over \(p\). A morphism \(f : \vartheta \to
  \psi\) consists in a morphism \(f : X \to Y\) such that, for any time \(t :
  \Tt\) and global section \(\sigma : p(1) \to \Sum_{i:p(1)} p[i]\) of \(p\), we
  have a natural transformation \(\vartheta^\sigma \to \psi^\sigma\) between the
  closures.  The identity morphism \(\id_\vartheta\) on the open Markov process
  \(\vartheta\) is given by the identity morphism \(\id_X\) on its state space
  \(X\). Composition of morphisms of open Markov processes is given by
  composition of the morphisms of the state spaces.
\end{prop}

\begin{prop} \label{prop:mproc-idx}
  \(\MProcT{\Tt}_{\Pa}\) extends to an indexed category, \(\MProcT{\Tt}_{\Pa} :
  \Poly{E} \to \Cat{Cat}\). Suppose \(\varphi : p \to q\) is a morphism of
  polynomials.  We define the functor \(\MProcT{\Tt}_{\Pa}(\varphi) : \MProcT{\Tt}_{\Pa}(p)
  \to \MProcT{\Tt}_{\Pa}(q)\) as follows.  Suppose \((X, \vartheta^o,
  \vartheta^u)\) is an object (open Markov process) in
  \(\MProcT{\Tt}_{\Pa}(p)\). Then \(\MProcT{\Tt}_{\Pa}(\varphi)(X, \vartheta^o,
  \vartheta^u)\) is defined as the triple \((X, \varphi_1 \circ \vartheta^o,
  \vartheta^u \circ {\vartheta^o}^\ast \varphi^\#) : \MProcT{\Tt}_{\Pa}(q)\). On
  morphisms, \(\MProcT{\Tt}_{\Pa}(\varphi)(f) : \MProcT{\Tt}_{\Pa}(\varphi)(X,
  \vartheta^o, \vartheta^u) \to \MProcT{\Tt}_{\Pa}(\varphi)(Y, \psi^o, \psi^u)\)
  is given by the same underlying map \(f : X \to Y\) of state spaces.
\end{prop}

\begin{rmk}[Closed Markov chains and Markov processes]
  A closed \textit{Markov chain} is given by a map \(X \to \Pa X\), where \(\Pa
  : \cat{E} \to \cat{E}\) is a probability monad on \(\cat{E}\); this is
  equivalently an object in \(\MProcT{\Tt}_{\Pa}(y)\), and (again equivalently)
  an object in \(\Cat{Cat}\big(\deloop{\nn}, \Kl(\Pa)\big)\). With more general
  time \(\Tt\), one obtains closed \textit{Markov processes}: objects in
  \(\Cat{Cat}\big(\deloop{\Tt}, \Kl(\Pa)\big)\). More explicitly, a closed
  Markov process is a time-indexed family of Markov kernels; that is, a morphism
  \(\vartheta : \Tt \times X \to \Pa X\) such that, for all times \(s,t : \Tt\),
  \(\vartheta_{s+t} = \vartheta_s \klcirc \vartheta_t\) as a morphism in
  \(\Kl(\Pa)\). Note that composition \(\klcirc\) in \(\Kl(\Pa)\) is given by
  the Chapman-Kolmogorov equation, so this means that
  \[
  \vartheta_{s+t}(y|x) = \int_{x':X} \vartheta_s(y|x') \, \vartheta_t(\d x'|x) \, .
  \]
\end{rmk}

We will need to convert our fibration of Markov processes into an ordinary
category, in order to supply `dynamical semantics' for statistical games. We can
do so, with objects being appropriately typed pairs and morphisms representing
appropriately `bidirectional' dynamical systems, by using the following
structure.

\begin{prop} \label{prop:hibi-d}
  Let \(\Fun{D} : \Poly{E} \to \Cat{Cat}\) be an indexed category over
  polynomials. Then there is a category of hierarchical bidirectional
  \(\Fun{D}\)-systems, denoted \(\HiBiD\), and defined as follows. The objects
  of \(\HiBiD\) are pairs \((X, A)\) of objects in \(\cat{E}\). Morphisms \((X,
  A) \to (Y, B)\) are functors \(\Fun{D}(Xy^A) \to \Fun{D}(Yy^B)\). The
  composition rule is just composition of the corresponding functors, and
  identity morphisms \(\id_{(X,A)} : (X, A) \to (X, A)\) are given by the
  identity functor \(\id_{\Fun{D}(Xy^A)} : \Fun{D}(Xy^A) \to \Fun{D}(Xy^A)\).
  \begin{proof}
    Immediate from the associativity and unitality of composition of functors.
  \end{proof}
\end{prop}

In particular, we can take \(\Fun{D} = \MProcT{\Tt}_{\Pa}\) to obtain
``hierarchical bidirectional Markov systems''.

\subsection{Nested Systems and Dependent Polynomials}

The foregoing formalism suffices to describe systems' shapes, and behaviours of
those shapes that depend on their sensoria. But in our world, a system has a
\emph{position} as well as a shape! Indeed, one might want to consider systems
nested within systems, such that the outer systems constitute the `universes' of
the inner systems; in this way, inner shapes may depend on outer shapes, and
inner sensoria on outer sensoria.\footnote{We might even consider the outer
shapes explicitly as positions in some world-space, and the outer sensorium as
determined by possible paths between positions, in agreement with the
perspective of \cite{Spivak2020Poly} on polynomials.} We can model this
situation polynomially.

Recall that an object in $\Poly{E}$ corresponds to a bundle $E \to B$,
equivalently a diagram $1 \leftarrow E \xto{p} B \to 1$, and note that the unit
polynomial $y$ corresponds to a bundle $1 \to 1$. We can then think of
$\Poly{E}$ as the category of ``polynomials in one variable'', or ``polynomials
over y''. This presents a natural generalization, to polynomials in many
variables, corresponding to diagrams $J \leftarrow E \to B \to I$; these
diagrams form the objects of a category $\Poly{E}(J, I)$. When $J$ is a
(polynomial) bundle $\beta$ over $I$, then we can take the subcategory of
$\Poly{E}(J, I)$ whose objects are commuting squares and whose morphisms are
prisms as follows; the commutativity ensures that inner and outer sensoria are
compatible.

\begin{prop}
  There is an indexed category of \textbf{nested polynomials} which by abuse of
  notation we will call $\Poly{E}(-) : \Poly{E} \to \Cat{Cat}$. Given $\beta : J
  \to I$, the category $\Poly{E}(\beta)$ has commuting squares as on the left
  below as objects and prisms as on the right as morphisms. Its action on
  polynomial morphisms $\beta \to \gamma$ is given by composition.
  
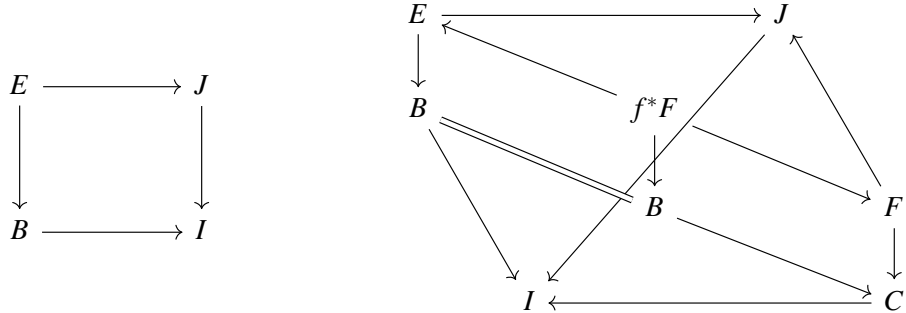
\begin{equation*}
    \begin{tikzcd}
      E && J \\
      \\
      B && I
      \arrow[from=1-3, to=3-3]
      \arrow[from=1-1, to=1-3]
      \arrow[from=1-1, to=3-1]
      \arrow[from=3-1, to=3-3]
    \end{tikzcd}
    \qquad\qquad\qquad
    \begin{tikzcd}
      E &&& J \\
      B && {f^*F} \\
      && B && F \\
      & I &&& C
      \arrow[from=1-4, to=4-2]
      \arrow[from=2-1, to=4-2]
      \arrow[Rightarrow, no head, from=2-1, to=3-3]
      \arrow[from=3-3, to=4-5]
      \arrow[from=2-3, to=3-5]
      \arrow[from=3-5, to=4-5]
      \arrow[from=2-3, to=1-1]
      \arrow[from=1-1, to=2-1]
      \arrow[from=4-5, to=4-2]
      \arrow[from=3-5, to=1-4]
      \arrow[from=1-1, to=1-4]
      \arrow[from=2-3, to=3-3]
    \end{tikzcd}
  \end{equation*}
\end{prop}

\begin{rmk}
  This construction can be repeatedly iterated, modelling systems within systems
  within systems. We leave the consideration of the structure of this iteration
  to future work, though we expect it to have an opetopic shape equivalent to
  that obtained by iterating the $\Cat{Para}$ construction (cf.~Remark
  \ref{rmk:para}).
\end{rmk}

\begin{obs}
  Our polynomially indexed categories of dynamical behaviours and statistical
  games (Prop.~\ref{prop:psgame}) generalize to the case of nested polynomials,
  giving a doubly-indexed structure. For more information on the former case, we
  refer the reader to our preprint \cite{Smithe2021Some}.
\end{obs}

\section{Theories of Approximate and Active Inference}

We now start to bring together the structures of the previous sections, in order
to breathe life into polynomials. We begin by sketching \emph{approximate
inference doctrines}, which characterize dynamical systems that optimize their
performance at statistical games, without reference to morphology. In this
paper, we do not concentrate on the detailed structure of these doctrines,
leaving their exposition and comparison to future work, in which we will also be
interested in morphisms between doctrines.

\subsection{Approximate Inference Doctrines}

An approximate inference doctrine will be a monoidal functor from a category of
statistical games into an appropriate category of dynamical systems, taking
games to systems that `play' those games, typically by implementing an
optimization process.  In the free-energy literature (\emph{e.g.},
\cite{Buckley2017free}), these systems have a hierarchical structure in which
the realization of a game has access to the dynamics realizing the prior,
mirroring the context-dependence of the games themselves, and it is for this
reason that we use the category of hierarchical bidirectional systems defined
above.

\begin{prop} \label{prop:SGame-F}
  Let \(\Pa : \cat{E} \to \cat{E}\) be a probability monad on \(\cat{E}\), and
  let \(\cat{C}\) be a category that admits Bayesian inversion, equipped with a
  functor \(F : \cat{C} \to \Kl(\Pa)\). Since \(\SGame{\cat{C}}\) is a
  Grothendieck fibration, it is equipped with a canonical projection functor
  \(\SGame{\cat{C}} \to \cat{C}\). Suppose that \(\cat{G}\) is a subcategory of
  \(\SGame{\cat{C}}\). Then there is a functor \(\Fun{F}_{\cat{G}} : \cat{G}
  \hookrightarrow \SGame{\cat{C}} \to \cat{C} \xto{F} \Kl(\Pa)\) whose image
  \(\im\Fun{F}_{\cat{G}}\) in \(\Kl(\Pa)\) consists of the forwards maps of
  \(\cat{G}\) realized as stochastic channels in \(\Kl(\Pa)\).
\end{prop}

\begin{defn}
  Suppose \(p : \Poly{E}\) and that \(\cat{C}\) is a subcategory of
  \(\Kl(\Pa)\). Denote by \(\MProcT{\Tt}_{\Pa}(p)|_{\cat{C}}\) the subcategory
  of \(\MProcT{\Tt}_{\Pa}(p)\) whose objects \((\Theta, \theta^o, \theta^u)\)
  satisfy the condition that, for all \(t : \mathbb{T}\), the following
  composite belongs to \(\cat{C}\):
  \[
  \Sum_{t:\mathbb{T}} \Sum_{x:\Theta} p[\theta^o(t, x)] \xto{\theta^u(t)} \Pa \Theta \xto{\Pa \theta^o(t)} \Pa p(1) \, .
  \]
  Note that \(\MProcT{\Tt}_{\Pa}(-)|_{\cat{C}}\) does not in general define an
  indexed category.
\end{defn}

\begin{defn}
  Suppose \(\cat{C}\) is a subcategory of \(\Kl(\Pa)\). Denote by
  \(\HiBiF{\MProcT{\Tt}_{\Pa}}|_{\cat{C}}\) the restriction of
  \(\HiBiF{\MProcT{\Tt}_{\Pa}}\) defined as follows. The objects of
  \(\HiBiF{\MProcT{\Tt}_{\Pa}}|_{\cat{C}}\) are the objects of
  \(\HiBiF{\MProcT{\Tt}_{\Pa}}\), but morphisms \((X,A) \to (Y,B)\) are now
  functors \(\MProcT{\Tt}_{\Pa}|_\cat{C}(Xy^A) \to
  \MProcT{\Tt}_{\Pa}|_\cat{C}(Yy^B)\). Composition and identities are defined as
  for \(\HiBiF{\MProcT{\Tt}_{\Pa}}\).
\end{defn}

We are now in a position to define approximate inference doctrines.

\begin{defn}
  Let \(\Pa : \cat{E} \to \cat{E}\) be a probability monad on \(\cat{E}\) and
  let \(\cat{G}\) be a subcategory of \(\SGame{\Kl(\Pa)}\). An
  \textbf{approximate inference doctrine} in \(\cat{G}\) with time \(\Tt\) is a
  functor from \(\cat{G}\) to
  \(\HiBiF{\MProcT{\Tt}_{\Pa}}|_{\im\Fun{F}_{\cat{G}}}\), where
  \(\Fun{F}_{\cat{G}}\) is defined as in Proposition \ref{prop:SGame-F}.
\end{defn}

Many informal approximate inference schemes---including Markov chain Monte
Carlo, variational Bayes, expectation-maximization, particle filtering---give
rise to approximate inference doctrines; functoriality typically follows from
Theorem $\ref{thm:buco}$. Here we note one explicitly, for later reference; a
detailed presentation will appear in an upcoming paper.

\begin{defn}
  We say that \(f : X \klto Y\) in \(\Kl(\Pa)\) is \textbf{Gaussian} if, for any
  \(x : X\), the state \(f(x) : \Pa Y\) is Gaussian. Note that Gaussian
  morphisms are not themselves closed under composition, though we can consider
  the subcategory of morphisms generated by Gaussian morphisms.
\end{defn}

\begin{lemma}[Laplace approximation] \label{lemma:laplace-approx}
  Suppose:
  \begin{enumerate}
  \item \(\Cat{Gauss}\) is the subcategory of channels generated by Gaussian
    morphisms between finite-dimensional Euclidean spaces in \(\Kl(\Pa)\);
  \item \((\gamma, \rho, \phi) : X \to Y\) is a simple \(D_{KL}\)-free energy
    game with Gaussian channels;
  \item for all priors \(\pi : \Pa X\), the statistical parameters of \(\rho_\pi : Y \to \Pa X\) are denoted by \((\mu_{\rho_\pi}, \Sigma_{\rho_\pi}) : Y \to \rr^{|X|} \times \rr^{|X|\times |X|}\), where \(|X|\) is the dimension of \(X\); and
  \item for all \(y : Y\), the eigenvalues of \(\Sigma_{\rho_\pi}(y)\) are small.
  \end{enumerate}
  Then the loss function \(\phi : \overline{\BLens{\Cat{Gauss}}}((X,X), (Y,Y))
  \to \rr\) can be approximated by
  \[
  \phi(\pi, k) = \E_{y \sim k \klcirc \gamma \klcirc \pi} \big[ \Fa(y) \big]
  \approx \E_{y \sim k \klcirc \gamma \klcirc \pi} \big[ \Fa^L(y) \big]
  \]
  where
  \begin{align} \label{eq:laplace-energy}
    \Fa^L(y)
    & = E_{(\pi,\gamma)}\left(\mu_{\rho_\pi}(y), y\right) - S_X \left[ \rho_\pi(y) \right] \\
    & = -\log p_\gamma(y|\mu_{\rho_\pi}(y)) -\log p_\pi(\mu_{\rho_\pi}(y)) - S_X \left[ \rho_\pi(y) \right] \nonumber
  \end{align}
  and where \(S_x[\rho_\pi(y)] = \E_{x \sim \rho_\pi(y)} [ -\log p_{\rho_\pi}(x|y)
  ]\) is the Shannon entropy of \(\rho_\pi(y)\), and \(p_\gamma : Y \times X \to
  [0,1]\), \(p_\pi : X \to [0,1]\), and \(p_{\rho_\pi} : X \times Y \to [0,1]\)
  are density functions for \(\gamma\), \(\pi\), and \(\rho_\pi\) respectively.
  The approximation is valid when \(\Sigma_{\rho_\pi}\) satisfies
  \begin{equation} \label{eq:laplace-sigma-rho-pi}
    \Sigma_{\rho_\pi} (y) = \left(\partial_x^2 E_{(\pi,\gamma)}\right)\left( \mu_{\rho_\pi}(y), y\right)^{-1} \, .
  \end{equation}
\end{lemma}

\begin{thm} \label{thm:laplace-doctrine}
  Suppose:
  \begin{enumerate}
  \item \(\Cat{Gauss}\) is the subcategory of channels generated by Gaussian
    morphisms between finite-dimensional Euclidean spaces in \(\Kl(\Pa)\);
  \item \(\SimpSGame{\Cat{Gauss}} \hookrightarrow \SGame{\Kl(\Pa)}\) is the
    subcategory of simple statistical games over \(\Kl(\Pa)\) with channels in
    \(\Cat{Gauss}\); and
  \item \(\cat{G}\) is the subcategory of \(D_{KL}\)-Bayesian inference games in
    \(\SimpSGame{\Cat{Gauss}}\).
  \end{enumerate}
  Then the discrete-time free-energy principle under the Laplace approximation
  induces an approximate inference doctrine in \(\cat{G}\) with time \(\nn\),
  \(\Fun{Laplace} : \cat{G} \to
  \HiBiF{\MProcT{\Tt}_{\Pa}}|_{\im\Fun{F}_{\cat{G}}}\).
\end{thm}

\subsection{Statistical Games over Polynomials}

Approximate inference doctrines of the foregoing type do not supply a
satisfactory model of active systems. One piece of structure is still missing,
with which we can describe action and interaction faithfully: an indexed
category of statistical games over polynomials. In order to construct this, we
first define categories of ``games on interfaces'': this is simpler than slicing
the category of statistical games, as we do not require games between games.

\begin{defn}
  Let $\Pa : \cat{E} \to \cat{E}$ be a probability monad on $\cat{E}$. Let $X :
  \cat{E}$ be an object in $\cat{E}$. Define a category of \textbf{simple
    statistical games on the interface} $X$, denoted by $\IntGame{\Pa}(X)$, as
  follows. Its objects are simple statistical games with codomain $X$; that is,
  points of $\sum_{A:\cat{E}} \SimpSGame{\Kl(\Pa)}(A, X)$. Let $(\gamma, \rho,
  \phi) : A \to X$ and $(\delta, \sigma, \chi) : B \to X$ be two such simple
  statistical games. Then a morphism $(\gamma, \rho, \phi) \to (\delta, \sigma,
  \chi)$ is a deterministic function $f : A \to B$---that is, a point of
  $\cat{E}(A, B)$---such that $\gamma = \delta \circ f$. Unitality and
  associativity follow immediately from those properties in $\cat{E}/\Pa X$.
\end{defn}

We then use this to construct games over polynomials. The intuition here is that
`inside' a system with a polynomial phenotype is a statistical model of the
system's sensorium. This involves an object representing the space of possible
causes of observations, and a simple statistical game from this object onto the
sensorium; by its nature, this model induces predictions about the system's
configurations, as well as about the immanent signals. By sampling from these
predicted configurations, the system can act; by observing its actual
configuration and the corresponding immanent signals, it can update its internal
beliefs, and any parameters of the model. Later, we will equip this process with
(random) dynamics, thereby giving the systems life.

\begin{prop} \label{prop:psgame}
  Let $\Pa : \cat{E} \to \cat{E}$ be a probability monad on a locally Cartesian
  closed category $\cat{E}$. There is a \textbf{polynomially indexed category of
    statistical games} $\PSGame{\Pa} : \Poly{E} \to \Cat{Cat}$, defined on
  objects $p$ as $\IntGame{\Pa}\left(\sum_{i:p(1)} p[i]\right)$.
\end{prop}

\begin{ex}
  To understand the action of $\PSGame{\Pa}(\varphi)$ on statistical games, it
  may help to consider the example of a corporation. Such a system is composed
  of a number of active systems, which instantiate statistical games and
  interact according to some pattern formalized by the polynomial map
  $\varphi$. Given such a collection of games, $\PSGame{\Pa}(\varphi)$ tells us
  how to construct a game for the corporation as a whole: in particular, we
  obtain a stochastic channel generating predictions for the (exposed) sensorium
  of the corporation, and an inversion updating the constituent systems' beliefs
  accordingly.
\end{ex}

\subsection{Active Inference}

We are now ready to define active inference doctrines; given all the foregoing
structure, this proves relatively simple.

\begin{defn}
  Let $\Pa : \cat{E} \to \cat{E}$ be a probablity monad, and let $\mathbb{T}$ be
  a time monoid. An \textbf{active inference doctrine} is a monoidal indexed
  functor from $\PSGame{\Pa}$ to $\MProcT{\Tt}_{\Pa}$.
\end{defn}

\begin{prop}
  The Laplace doctrine lifts from approximate to active inference. The functors
  on each fibre are as before on games (here, objects), and on morphisms between
  games they are given merely by lifting the corresponding maps to maps between
  state spaces. One then checks that morphisms of polynomials correspond to
  natural transformations between these functors.
\end{prop}

\begin{rmk}
  It would be desirable to incorporate the compositional structure of the games
  themselves, rather than treat them opaquely as objects. This suggests a
  double-categorical structure the investigation of which we leave to future
  work. Similarly, we do not here elaborate the extension of these indexed
  categories to the dependent-polynomial case.
\end{rmk}

\section{Polynomial Life and Embodied Cognition}

Finally, we sketch how a number of classic biological processes can be modelled
as processes of active inference over polynomials. The key insight is that, by
fixing the prior of an `active' free-energy game to encode high-precision
(low-variance) beliefs about the external state, we can induce the system to
prefer acting (\emph{i.e.}, reifying those beliefs) over perceiving
(\emph{i.e.}, updating the beliefs to match perceptions). In doing so, one can
induce \emph{volition} or \emph{goal-directedness} in the system. A key feature
of these examples is that they demonstrate `embodied' cognition, in which a
system's form and interactions become part of its cognitive apparatus.

\begin{rmk}
  Of course, one must be careful not to choose a prior with excessively high
  precision (such as a Dirac delta distribution), as this would cause the system
  to forego any belief-updating, thereby rendering its actions independent of
  the `actual' external state.
\end{rmk}

\begin{ex}
  Suppose that the system's sensorium includes a key parameter such as ambient
  temperature or blood pH\@. Suppose that by adjusting its configuration, the
  system can move around in order to sample this parameter. And suppose that the
  prior encodes a high-precision distribution centred on the acceptable range
  of this parameter.  Then it is straightforward to show that the system, by
  minimizing the free energy, will attempt to configure itself so as to remain
  within the acceptable parameter range. We can consider this as a simple model
  of \textbf{homeostasis}.
\end{ex}

\begin{ex}
  We can extend the previous example to a system with multiple (polynomial)
  components, each equipped with a ``homeostasis game'', in order to model
  \textbf{morphogenesis}. Suppose the environmental parameter in the sensorium
  is the local concentration of some signalling molecule, and suppose the
  polynomial morphism forming the composite system encodes the pattern of
  signalling molecule concentrations in the neighbourhood of each system, as a
  result of their mutual configurations. Suppose then that the target state
  encoded in the prior of each system corresponds to the system being positioned
  in a particular way relative to the systems around it, as represented by the
  signal concentrations. Free-energy minimization then induces the systems to
  arrange themselves in order to obtain the target pattern.
\end{ex}

\begin{rmk}
  The foregoing examples begin to point towards a compositional theory of
  \emph{autopoiesis}: here, one might expect the target state to encode the
  proposition ``maintain my morphology'', which appears self-referential. The
  most elegant way of encoding this proposition in the prior is not immediately
  clear, although a number of possibilities present themselves (such as avoiding
  some undesirable configuration representing dissolution). We expect a
  satisfactory answer to this to be related to ``Bayesian mechanics''
  (see \secref{sec:future}).
\end{rmk}

\begin{rmk}
  It has been shown informally that, given a finite time horizon Markov decision
  problem, active inference can recover the Bellman-optimal policy traditionally
  obtained by backward induction \cite{DaCosta2020Relationship}. Ongoing work
  by the present author is directed at formalizing the structure of this
  relationship. In particular, the result rests on encoding directly into the
  prior the expectation of the loss function given a policy and a goal, which
  strikes us as a large amount of information to push into an unstructured
  distribution over numbers.
\end{rmk}

\begin{ex}
  The examples need not be restricted to simple biological cases. For instance,
  we can model spatial navigation quite generally: we can use parameterized
  statistical games to encode uncertainty about the structure of the `external
  space' (for instance: which points or neighbourhoods are connected to which,
  and by which paths). By setting a high-precision prior at some location, the
  system will attempt to reach that location, learning the spatial structure
  along the way; reducing the precision of the prior causes the system to prefer
  ``mere exploration''. One can attach sense-data to each location using the
  natural polynomial bundle structure. Moreover, the `external space' need not
  be a simple topological space: it may be something more structured. For
  instance, categories and sites can themselves be modelled polynomially. One
  can think of ``taking an action'' as precisely analogous to ``following a
  morphism'': thus, in a topos-theoretic setting, one can consider the structure
  of the `external space' to be a type-theoretic context, and positions in the
  world to be objects in the corresponding topos. One could then encode in the
  prior a target proposition, and free-energy minimization would cause the
  system then to explore the `space' (learning its structure), and seek a path
  to the target. But such a path is precisely a proof! There is increasing
  evidence that the neural mechanisms underlying spatial and abstract navigation
  are the same \cite{Behrens2018What}, and a line of thinking as sketched here
  may supply a mathematical justification.
\end{ex}

\section{Future Directions} \label{sec:future}

Besides expanding the examples above in detail, there are many future directions
to pursue. Our last example points towards a `well-typed' theory of cognition,
finding type-theoretic analogues of cognitive processes (such as action,
planning, or navigation). By formalizing the connection between polynomial
statistical games and Markov decision processes, we hope finally to relate our
`statistical' account of cybernetics with the account emerging from research in
compositional game theory. In particular, we believe that the
hierarchical/nested structure of our polynomial systems is structurally similar
to that of (parameterized) players in open games. Along similar lines, we expect
a connection between statistical games and `learners' \cite{Spivak2021Learners}
implementing backprop.

In a more physical direction, there is a controversy in the informal literature
about whether one should expect \emph{any} system with a boundary (and hence any
system on a polynomial) to admit a canonical statistical-game description; the
typical suggestion is that such a description should obtain at non-equilibrium
steady state, through a manipulation of the corresponding Fokker-Planck
equation; this is the notion of ``Bayesian mechanics''
\cite{Friston2019free}. Our results suggest that such a canonical description
should form a left adjoint to some active inference doctrine; this is a matter
of ongoing research by the author.

Working topos-theoretically points further in a metaphysical direction: a
Bayesian perspective lends itself to subjectivism, but considering the
``internal universe'' of a navigating system to be a topos in some context also
points to a subjective realism. It seems likely then that composite systems need
not in general agree about their observations. We should therefore expect to
find evidence of contextuality and disagreement in multi-agent systems, and to
investigate this using cohomological tools (\emph{e.g.},
\cite{Caru2017Cohomology}).

\bibliographystyle{eptcs}
\bibliography{bibliography}

\end{document}